\documentclass[aps,twocolumn,nofootinbib]{revtex4-2}

\usepackage{bbold}
\usepackage{amsmath}
\usepackage{amsfonts}
\usepackage{amssymb}

\def\[{\left\lbrack}
\def\]{\right\rbrack}

\def\({\left(}
\def\){\right)}
\newcommand{\be}{\begin{equation}}
\newcommand{\ee}{\end{equation}}
\newcommand{\ea}{\end{eqnarray}}
\newcommand{\ba}{\begin{eqnarray}}

\usepackage{colordvi}
\usepackage{color}

\def\I{\mathbb{1}}
\def\O{\mathbb{0}}
\newcommand{\q}{\mathbf{q}}
\newcommand{\p}{\mathbf{p}}

\renewcommand{\o}{\mathbf{0}}

\begin{document}

\title{Barcelos-Wotzasek symplectic algorithm for constrained systems revisited}

\author{M.A. de Andrade}\thanks{e-mail: marco@fat.uerj.br}
\author{C. Neves}\thanks{e-mail: clifford@fat.uerj.br}
\author{E.V. Corr\^{e}a Silva}\thanks{e-mail: eduardo.vasquez@fat.uerj.br}
\affiliation{Departamento de Matem\'{a}tica, F\'\i sica e Computa\c{c}\~{a}o, \\
Faculdade de Tecnologia, \\ 
Universidade do Estado do Rio de Janeiro,\\
Rodovia Presidente Dutra, Km 298, P\'{o}lo
Industrial,\\
CEP 27537-000, Resende-RJ, Brazil.}
\begin{abstract}
	A minor change in the Barcelos-Wotzasek (BW) symplectic algorithm for
	constrained systems is proposed. The change addresses some criticism
	that formalism has received, placing it on the same footing as Dirac's
	algorithm.

	\end{abstract}
\maketitle
\setlength{\baselineskip}{20 pt}

\section{Introduction}
\label{intro}

Constrained models can usually be handled by Dirac's method
\cite{dirac1,dirac2,dirac3,dirac4}, in which the Hamiltonian plays the central role.
Some alternative ways, however, develop from the Lagrangian
instead \cite{BW1,BW2,Shirzad1,rothe1,Shirzad2,rothe2}. The
Barcelos-Wotzasek (BW) symplectic algorithm \cite{BW1,BW2} is one
example which has been widely and properly
applied to a multiplicity of models
\cite{Anjali1,Omar,Nahomi,Anjali2,Davi,Escalante,Manavella,Huang,Cheb-Terrab}, despite some criticism on its limitations
\cite{rothe1,Shirzad2,rothe2}, e.g., the primary constraints set is not taken into account when obtaining the first-order Lagrangian, the Dirac's set of constraints are not restored, the failure of symplectic analysis for a first or second-class system with only one primary constraint, or fail whenever second-class constraints emerge at third level or higher.

The purpose of this article is that of showing how the aforementioned
criticisms can be circumvented by a minor extension of a fundamental
concept in the BW formalism: namely, that {\it all} constraints must be
introduced into the kinetic part of the Lagrangian, rather than set
strongly to zero in the symplectic potential. With that minor but
significant \textit{ansatz}, the BW formalism can be regarded on an
equal foot with Dirac's method, since the results provide by the Dirac's method are reproduced. However, it is important to notice that we do not provide a mathematical formal proof of equivalence to the Dirac's method, since a connection between the our proposal symplectic structure and pre-symplectic structure of constrained dynamics is not establish, as done in Ref.\cite{gotay}.

Our article is organized as follows. Section \ref{ComparisonDiracBW}
compares how Dirac's and BW methods are applied to three constrained
models, illustrating the criticisms made to the BW method, regarding its
discrepancies with respect to that of Dirac's: the free particle in a
hypersphere (subsection \ref{HypersphereDiracBW}); a toy model
investigated in Ref.\cite{rothe1,rothe2} (subsection \ref{ToyDiracBW});
and the relativistic free particle (subsection
\ref{RelativisticDiracBW}). In Section \ref{BW+}, we state our extension
of the BW method, and subsequently apply it to the aforementioned
models, showing that all discrepancies have been eliminated: the free
particle in a hypersphere (subsection \ref{HypersphereBW+}); the toy
model (subsection \ref{ToyBW+}); and the relativistic free particle
(subsection \ref{RelativisticBW+}). Section \ref{conclusion} contains
our concluding remarks.

\section{Comparison of Dirac's and Barcelos-Wotzasek algorithms}
\label{ComparisonDiracBW}

Before we detail our proposal of extension to the BW method in section
\ref{BW+}, we shall motivate it and put it in the correct perspective, by studying three
models from both Dirac's \cite{dirac1,dirac2,dirac3} and Barcelos-Wotzasek (BW) \cite{BW1,BW2} perspectives: the
free particle on a hypersphere, a toy model, and the relativistic free
particle.

\subsection{Free Particle on a Hypersphere}
\label{HypersphereDiracBW}

\subsubsection{From Dirac's perspective}
\label{sec11}

Consider a free particle on a $(N-1)$-hypersphere, with its dynamics governed by the following Lagrangian,
\be 
\label{0000}
{\cal L}_{\text{hyp}}(q,\dot q, \lambda)=\frac 1 2 \sum_{i=1}^N \dot q_i^2 + \frac {\lambda}{2}\left(\sum_{i=1}^Nq_i^2-1\right),
\ee 
in which $\lambda$ is a Lagrange multiplier, and
$q\equiv[q_1,\ldots,q_N]$ are the particle coordinates in the
$N$-dimensional space. This Lagrangian is defined in the expanded
coordinate-velocity space, which includes the multiplier,
$\left[q_i,\dot q_i,\lambda\right]$, in which $i\in\{1,~2,\dots ,~N\}$.

This model has the primary constraint
\be
\label{0010}
\phi=\pi,
\ee
in which $\pi$ is the canonical conjugate momentum  to $\lambda$. From the Lagrangian (\ref{0000}) one obtains the Hamiltonian
\be
\label{0020}
{\cal H}_{\text{hyp}}(q,p,\lambda)=\frac 12 \sum_{i=1}^N p_i^2 - \frac {\lambda}{2}\left(\sum_{i=1}^Nq_i^2-1\right),
\ee
in which $p_i$ is the canonical conjugate momentum to $q_i$, for $i\in\{1,~2,\dots ,~N\}$. 

Dirac's procedure yields the following set of secondary constraints,
\ba
\label{0030}
\Omega_1&=&\frac {1}{2}\left(\sum_{i=1}^Nq_i^2-1\right),\nonumber\\
\Omega_2&=&\sum_{i=1}^Nq_i p_i,\\
\Omega_3&=& \sum_{i=1}^N (p_i^2 +\lambda q_i^2),\nonumber
\ea
and the following Dirac's brackets,
\ba 
\label{0040}
\lbrace q_i, q_j\rbrace &=& 0,\nonumber\\
\lbrace q_i, p_j\rbrace&=& \delta_{ij}-\frac{q_iq_j}{q^2},\\
\lbrace p_i, p_j\rbrace&=& \frac{p_iq_j-q_ip_j}{q^2},\nonumber
\ea 
in which $i, j\in\{1,~2,\dots ,~N\}$, $q^2=\sum_{i=1}^Nq_i^2.$ 
The Hamiltonian assumes the form
\be
\label{0020a}
{\cal H}_{\text{hyp}}(q,p,\lambda)=\frac 12 \sum_{i=1}^N p_i^2.
\ee

\subsubsection{From the BW perspective}
\label{sec21}

The second-order Lagrangian (\ref{0000}) of subsection \ref{sec11} can be rewritten in first-order form as
\be 
\label{0100}
{\cal L}_{\text{hyp}}(q,\dot q,p, \lambda)= \sum_{i=1}^{N}p_i\dot q_i - {\cal H}_{\text{hyp}}(q,p,\lambda)        ,
\ee
in which the symplectic potential ${\cal H}_{\text{hyp}}$ is given by (\ref{0020}) and the
symplectic variables form the vector
\be\label{0100a}
\xi=\left(
\begin{array}{c}
\q     \cr
\p     \cr
\lambda 
\end{array}\right),
\ee
in which $\q$, $\p$ are $N\times1$ vectors whose components are respectively $\left[q_1, q_2, \dots, q_N\right]$ and $\left[p_1, p_2, \dots, p_N\right]$.
According to the BW formalism, the model yields the second-order tensor
\be
\label{0110}
f_{\text{hyp}}=\left(
\begin{array}{ccc}
\O    ~&~  -\I    ~&~ \o \cr
\I    ~&~  \O     ~&~ \o \cr
\o^T ~&~  \o^T    ~&~ 0
\end{array}\right),
\ee
\noindent 
in which $\O$ and $\I$ are the $N\times{}N$ null and $N\times{}N$ identity matrices, respectively; 
$\o$ is the $N\times1$ null vector and $\o^T$ is its transpose.
The $f_{\text{hyp}}$ matrix is singular; consequently, it has a zero-mode,
\be
\nu=\left(
\begin{array}{c}
\o \cr
\o \cr
1  
\end{array}\right).
\ee
(The arrangement of lines and columns of $f_{\text{hyp}}$ and $\nu$ reflects that of
variables in (\ref{0100a}), naturally.)

From the contraction of this zero-mode with the gradient of the symplectic potential ${\cal H}_{\text{hyp}}$, the following constraint is obtained,
\be
\label{0120} 
\Omega_1=\nu^T \, \frac{\partial {\cal H}_{\text{hyp}}}{\partial\xi}=\frac 12\left(\sum_{i=1}^{N}q_i^2-1\right),
\ee
which is then introduced into the kinetic sector of the first-order Lagrangian (\ref{0100}), together with a Lagrange multiplier $\eta_1$, thus yielding
\be
\label{0130}
{\cal L}_{\text{hyp}}^{(1)}(q, \dot q, p, \dot \eta_1)= \sum_{i=1}^{N} p_i\dot q_i +\Omega_1\dot \eta_1 - {\cal H}_{\text{hyp}}^{(1)}(q,p),
\ee 
in which
\be
{\cal H}_{\text{hyp}}^{(1)}(q,p) = \left.{\cal H}_{\text{hyp}}(q,p,\lambda)\rule{0mm}{4mm}\right|_{\Omega_1=0}.
\ee
The constraint $\Omega_1$ is set strongly to zero in the symplectic potential ${\cal H}_{\text{hyp}}$. Therefore, ${\cal H}_{\text{hyp}}^{(1)}$ and thus ${\cal L}_{\text{hyp}}^{(1)}$ do not depend on $\lambda$.
The symplectic variables now form the vector
\be
\xi^{(1)}=\left(
\begin{array}{c}
\q     \cr
\p     \cr
\eta_1 
\end{array}\right),
\ee
and one obtains
\be
\label{0140}
f_{\text{hyp}}^{(1)}=\left(
\begin{array}{ccc}
\O    ~&~  -\I  ~&~ ~\q \cr
\I    ~&~  \O   ~&~ ~\o \cr
-\q^T ~&~  \o^T ~&~ ~0
\end{array}\right),
\ee
in which $\q^T$ is the transpose of $\q$.
This matrix has the zero-mode
\be
\nu^{(1)}=\left(
\begin{array}{c}
\o \cr
\q \cr
1  
\end{array}\right),
\ee
which, in its turn, generates the constraint
\be 
\label{0150}
\Omega_2=\nu^{(1)T} \, \frac{\partial {\cal H}_{\text{hyp}}^{(1)}}{\partial\xi^{(1)}}=\sum_{i=1}^{N} q_ip_i.
\ee
The new first-order Lagrangian reads
\be
\label{0160}
{\cal L}_{\text{hyp}}^{(2)}(q,\dot q,p, \dot\eta_1,\dot\eta_2)= \sum_{i=1}^{N}p_i\dot q_i +\Omega_1\dot \eta_1 +\Omega_2\dot \eta_2- {\cal H}_{\text{hyp}}^{(2)}(q,p),
\ee 
in which
\be
{\cal H}_{\text{hyp}}^{(2)}(q,p) = \left.{\cal H}_{\text{hyp}}^{(1)}(q,p)\rule{0mm}{4mm}\right|_{\Omega_2=0}.
\ee
The new symplectic variables form the vector
\be
\xi^{(2)}=\left(
\begin{array}{c}
\q     \cr
\p     \cr
\eta_1 \cr
\eta_2
\end{array}\right),
\ee
and one obtains
\be
\label{0170}
f_{\text{hyp}}^{(2)}=\left(
\begin{array}{cccc}
\O            ~&~     -\I       ~&~ \q    ~&~     ~\p  \cr
\I            ~&~     \O        ~&~ \o    ~&~     ~\q  \cr
-\q^T         ~&~     \o^T      ~&~ 0     ~&~     ~0   \cr
-\p^T         ~&~    -\q^T      ~&~ 0     ~&~     ~0
\end{array}\right),
\ee
%
%
%
in which $\p=[p_1, p_2, \dots, p_N]$ and $\p^T$ is its transpose.

The matrix $f_{\text{hyp}}^{(2)}$ is nonsingular; its inverse yields the Dirac brackets given by Eq.(\ref{0040}). 
Despite some criticism \cite{rothe1,Shirzad2,rothe2} on the BW procedure, we are led to the {\it same} results
obtained by these authors and as well as by using Dirac's method, as shown in the subsection~\ref{sec11}. At this point, we would like to stress that these BW formalism criticism does not proceeds since in this formalism the constraints do not appear and are not classified as in the Dirac's sense. This is because the restrictions are inserted in the Lagrangian kinetic sector. 

\subsection{A Toy Model}
\label{ToyDiracBW}

\subsubsection{From Dirac's Perspective}
\label{sec12}

Consider now a system whose dynamics is governed by the Lagrangian
\cite{rothe2}
\be 
\label{0050}
{\cal L}_{\text{toy}}(x,y,\dot x,\dot y)=\frac 12 \dot x^2-a\;x\dot y + \frac b2 (x-y)^2,
\ee
from which one obtains the canonical momenta,
\ba{\label{0060}}
p_x &=& \dot x,\nonumber\\
p_y&=& -ax,
\ea
conjugate to $x$ and $y$, respectively. Eq.(\ref{0060}) yields the primary constraint
\be
\label{0070}
\varphi=p_y + ax.
\ee
The canonical Hamiltonian, corresponding to the Lagrangian (\ref{0050}), is
\be 
\label{0075}
{\cal H}_{\text{toy}}(x,y,p_x,p_y)= \frac{p_x^2}{2} -\frac b2 (x-y)^2.
\ee
By following Dirac's procedure, the set of constraints
\ba 
\label{0080}
\Theta_1&=& ap_x-b(x-y),\nonumber\\
\Theta_2&=& -bp_x+ab(x-y),
\ea 
is obtained. The Dirac matrix of Poisson brackets among the constraints then takes the form
\be
\label{0090}
C = \left(
\begin{array}{ccc}
0 & (-b+a^2) & 0\\
-(-b+a^2) & 0 & (b^2-a^2b)\\
0 & -(b^2-a^2b) & 0
\end{array}
\right).
\ee
This is a singular matrix, regardless of the values assigned to $a$ and
$b$; consequently, there is a symmetry and it is not possible to get the
Dirac's brackets among the phase space variables. At this point, it is
important to notice that this result, obtained by a straightforward
computation and based on the well known Dirac's method, is {\it not}
the same obtained in Ref.\cite{rothe2}, where another formalism was
applied\footnote{\textit{Cf.} page 64 of Ref.\cite{rothe2}.}. 

The set of constraints $\{\varphi,~\Theta_1,~\Theta_2\}$ can be split into two subsets, 
$\{\varphi,~\Theta_1\}$ and $\{\Theta_2\}$. Note that 
$\varphi$ and $\Theta_1$ are second-class constraints and, consequently, their
partial Dirac's brackets can be computed by taking the submatrix of (\ref{0090}),
\be 
\label{0092}
C_s = \left(
\begin{array}{cc}
0 & (-b+a^2) \\
-(-b+a^2) & 0 
\end{array}
\right).
\ee
From $C_s^{-1},$ following Dirac's procedure, one obtains the following non-null partial Dirac brackets among the phase space variables,
\ba 
\label{0093}
\lbrace x,y\rbrace&=&\frac{a}{a^2-b},\nonumber\\
\lbrace x,p_x\rbrace&=&\frac{-b}{a^2-b},\nonumber\\
\lbrace y,p_x\rbrace&=& -\frac{b}{a^2-b},\nonumber \\
\lbrace y,p_y\rbrace&=&\frac{a^2}{a^2-b}.
\ea 
At this point, $\varphi$ and $\Theta_1$ are strongly equal to zero and, consequently, the
Hamiltonian (\ref{0075}) reduces to
\be 
\label{0094}
{\cal H}_{\text{toy}}(x,y,p_x,p_y)=\frac{(b-a^2)}{2b}p_x^2.
\ee

In addition, the time derivative of $\Theta_2$ does not generate a new
constraint. Hence, $\Theta_2$ is the only constraint left. It is a
first-class one, and it is also the generator of the following
infinitesimal transformations,
\ba 
\label{0095}
\delta x&=& -b\varepsilon,\nonumber\\
\delta p_x&=& 0,\\
\delta y&=&-b\varepsilon,\nonumber
\ea
which leave the Hamiltonians (\ref{0075}) and (\ref{0094}) invariant $(\delta{\cal H}_{\text{toy}} = 0)$.
It is important to note that when $b=a^2$, the matrix $C$ (\ref{0090}) is null, and so is the Hamiltonian.

\subsubsection{From the BW perspective}
\label{sec22}

The second-order Lagrangian (\ref{0050}) is rewritten in its first-order form as
\be 
\label{0180}
{\cal L}_{\text{toy}}(x,y,\dot x,\dot y,p_x)=p_x \dot x-ax\dot y - {\cal H}_{\text{toy}},
\ee
in which
\be 
\label{0190}
{\cal H}_{\text{toy}}(x,y,p_x,p_y)= \frac{p_x^2}{2} - \frac b2 (x-y)^2.
\ee 
The symplectic variables are 
\be 
\xi=\[x,p_x,y\]
\ee
and the corresponding $f$ matrix is
\be
\label{0200}
f_{\text{toy}}=\left( 
\begin{array}{ccc}
0&-1&-a\\
+1&0&0\\
+a&0&0
\end{array}
\right),
\ee
which has the zero-mode
\be 
\label{0210}
\nu=\left(
\begin{array}{c}
0 \cr a \cr -1
\end{array}
\right).
\ee 
In accordance with the BW formalism, we get the constraint
\be 
\label{0220}
\Theta_1=\nu^T\, \frac{\partial {\cal H}_{\text{toy}}}{\partial \xi}=ap_x-b(x-y),
\ee
which is subsequently introduced into the kinetic sector of the first-order Lagrangian (\ref{0180}), yielding
\be 
\label{0230}
{\cal L}_{\text{toy}}^{(1)}(x,y,\dot x,\dot y,p_x, \dot\lambda_1)=p_x \dot x-a\;x\dot y + \Theta_1\dot\lambda_1- {\cal H}_{\text{toy}}^{(1)},
\ee
in which
\be
\label{0235}
{\cal H}_{\text{toy}}^{(1)} = \left.{\cal H}_{\text{toy}}\rule{0mm}{4mm}\right|_{\Theta_1=0}=\frac{(b-a^2)p_x^2}{2b}.
\ee
Such procedure is repeated, with the new symplectic variables
\be
\xi^{(1)}=\[x,p_x,y,\lambda_1\],
\ee
yielding
\be
\label{0250}
f_{\text{toy}}^{(1)}=\left( 
\begin{array}{cccc}
0&-1&-a&-b\\
+1&0&0&+a\\
+a&0&0&+b\\
+b&-a&-b&0
\end{array}
\right),
\ee
the determinant of which is
\be 
\label{0260}
\det(f_{\text{toy}}^{(1)})=(b-a^2)^2.
\ee
If $\det\!\left(f_{\text{toy}}^{(1)}\right)= 0$, e.g., $b=a^2$, the results obtained in subsection
\ref{sec12} are reproduced. If $\det\!\left(f_{\text{toy}}^{(1)}\right)\neq 0$ then $f_{\text{toy}}^{(1)}$
is invertible, and
\be
\label{0270}
{f_{\text{toy}}^{(1)}}^{-1}= \frac{1}{(a^2-b)} \left( 
\begin{array}{cccc}
0&-b&a&0\\
b&0&b&-a\\
-a&-b&0&+1\\
0&+a&-1&0
\end{array}
\right).
\ee
The existence of ${f_{\text{toy}}^{(1)}}^{-1}$ implies that there are no symmetries in the model,
and its elements correspond to the Dirac brackets among the symplectic variables. 
Hamilton's equations are thus obtained as
\ba 
\label{0290} 
\dot x&=&\left\{ x,{\cal H}_{\text{toy}}^{(1)}\right\}=p_x,\nonumber\\
\dot p_x&=&\left\{ p_x,{\cal H}_{\text{toy}}^{(1)}\right\}=0,\nonumber\\
\dot y&=&\left\{ y,{\cal H}_{\text{toy}}^{(1)}\right\}=p_x,\\
\dot \lambda_1&=&\left\{ \lambda_1,{\cal H}_{\text{toy}}^{(1)}\right\}=-\frac{a}{b}p_x.\nonumber
\ea

At this point, it is important to notice that the BW algorithm conflicts
with Dirac's method (\textit{cf.} subsection \ref{sec12}). 
The time derivative of the third equation in (\ref{0290}) is
\be 
\label{0300}
\ddot y = 0 \Rightarrow \dot y= c,
\ee 
in which $c$ is constant.
From (\ref{0300}) and the third equation of (\ref{0290}), one obtains {\it yet another constraint},
\be 
\label{0310}
\Sigma = p_x - c,
\ee
which is {\em not} generated by BW algorithm. 

Unlike the case of the free particle on the hypersphere (cf. section \ref{HypersphereDiracBW}), the results 
provided by the BW formalism for the toy model (\ref{0050}) {\it conflict} with those obtained by
Dirac's method in subsection~\ref{sec12}.

\subsection{The Relativistic Free Particle}
\label{RelativisticDiracBW}

\subsubsection{From Dirac's Perspective}
\label{RFP}

Consider now the following action \cite{Marnelius} for the relativistic free particle,
\be 
\label{00010}
S = -mc^2\int d\tau,
\ee 
in which $\tau$ is the proper time of the particle. The infinitesimal world line length is
\ba
ds &=&c\;d\tau=(dx^\mu dx_\mu)^{1/2}\nonumber\\
&=& (\dot x^\mu \dot x_\mu)^{1/2} dt,
\ea
in which $x^\mu\equiv x^\mu(t)$ are the space-time coordinates, $t$ is
an arbitrary parameter along the world line, $\dot x^\mu=dx^\mu/dt$, and
the space-time metric is $(+,-,-,-)$. The action (\ref{00010}) can be rewritten as
\be 
\label{00030}
S = -mc\int ds = -mc \int dt\ (\dot x^\mu \dot x_\mu)^{1/2}.
\ee
Note that the action (\ref{00010}) is invariant under the reparametrization $\tau^\prime=f(\tau)$. The Lagrangian is
\be 
\label{00040}
{\cal{L}}=-mc(\dot x^\mu \dot x_\mu)^{1/2},
\ee
and the conjugate canonical momentum to $x^\mu$ is 
\be 
\label{00050}
p^\mu = -\frac{\partial{\cal{L}} }{\partial \dot x_\mu}=\frac{mc\dot x^\mu}{(\dot x^\nu \dot x_\nu)^{1/2}}.
\ee
The Poisson brackets can be naively computed as
\be 
\label{00060}
\left\{p^\nu,x_\mu\right\}=  \delta^\nu_\mu,~~\left\{p^\nu,p_\mu\right\}= \left\{x^\nu,x_\mu\right\}=0.
\ee
However, as it is well known, the determinant of the Hessian matrix is null; therefore, the relativistic free particle is a constrained system.
The constraint arises, within the Lagrangian framework, precisely from the canonical momentum (\ref{00050}),
\ba  
p^\mu p_\mu&=&\frac{mc\dot x^\mu}{(\dot x^\nu \dot x_\nu)^{1/2}}p_\mu = m^2c^2\ ,
\ea
which yields the constraint
\be 
\label{00080}
\phi=p^\mu p_\mu-m^2c^2.
\ee
This constraint can be written in equivalent form, given by
\be 
\label{00080a}
\sqrt{p_i^2 +m^2}-|p_0|=0.
\ee
According to Dirac's method, this is the only constraint of the theory,
since $\dot\phi=\left\{\phi,{\cal{H}}\right\}=0$. Then, $\phi$ is a
first-class constraint and, also, the generator of the infinitesimal
transformation
\ba 
\label{00090}
\delta x^\mu&=&2p^\mu,\nonumber\\
\delta p^\mu&=&0.
\ea

The Lagrangian (\ref{00040}) might be written in the phase-space coordinates $(x_\mu,p_\mu)$, reading
\ba  
\label{00041}
{\cal{L}}&=&-p^\mu \dot x_\mu,\nonumber\\
{\cal{L}}+p^\mu \dot x_\mu&=&0.
\ea
As well discussed in Ref.\cite{gitman1}, the gauge fixing condition is $x_0=\zeta ct$, with $\zeta=-sign  ~p_0$, which drives (\ref{00041}) to be
\ba
{\cal{L}}+ p^0\dot x_0+p^i\dot x_i&=&0, ~i\in \{1,2,3\},\nonumber\\
{\cal{L}}+\zeta p^0c+p^i\dot x_i&=&0,\nonumber\\
\zeta p^0c&=&p_i\dot x_i-{\cal{L}}, \nonumber\\
{\cal{H}}&=&p_i\dot x_i-{\cal{L}}, \nonumber
\ea
which is the Legendre transformation. The Hamiltonian is
\be 
\label{00070}
{\cal{H}}=\zeta p^0c.
\ee
Then, $\zeta p^0=E/c$, since ${\cal{H}}$ is the energy. Note that a new
constraint $(\Sigma=x_0-\zeta ct)$ has been imposed and, from Dirac's point of view, it fixes the symmetry. The variable $\zeta$ is not fixed by constraints and should be considered as an equal rights dynamical variable and assumes two values $\zeta=\pm 1$. For $\zeta=1$
the (nonsingular) Dirac's matrix can thus be computed, yielding the non-vanishing Dirac's brackets among phase-space coordinates,
\be 
\lbrace p^\nu,x_\mu\rbrace= \delta^\nu_\mu-\delta^\nu_0\frac{p_\mu}{p^0}.
\ee

\subsubsection{From the BW perspective}
\label{RFP-BW}

The first-order Lagrangian (\ref{00040}) is
\be 
\label{000100}
{\cal{L}}= - p^\mu\dot x_\mu.
\ee
Again, from a naive point of view, the symplectic variables might be assumed to be 
$\xi^\alpha=(x^\mu, p^\mu)$ and the corresponding symplectic matrix to be
\be 
\label{000110}
f=\left(
\begin{array}{cc}
0 & \delta^\mu_\nu\\
-\delta^\mu_\nu & 0
\end{array}
\right).
\ee
This matrix is nonsingular and its inverse generates the Poisson's
brackets given in Eq.(\ref{00060}). 

Note that this approach conflicts with the investigation carried out in
subsection \ref{RFP}. The constraint (\ref{00080}), {\it does not}
appear in the symplectic approach: it arises from the Lagrangian
framework, as shown before. In Dirac's method, this type of constraint
gets inserted into the Hamiltonian through Lagrange multipliers. This
procedure is not possible, though, within the symplectic formalism: it
would lead us away from a first-order Lagrangian such as (\ref{000100}). However, in Ref.\cite{Gavrilov} this kind of problem was solve by generalizing the Faddeev-Jackiw symplectic approach to non-autonomous constrained systems.

\section{Barcelos-Wotzasek symplectic algorithm modified}
\label{sec3}
\label{BW+}

In order to handle the kind of problem manifested in subsections \ref{ToyDiracBW} and \ref{RelativisticDiracBW},
we propose minor but significant adjustments to the BW
symplectic algorithm that put the latter on an equal footing with Dirac's
method. 
\begin{itemize}
\item {\it All} types of constraints --- 
	whatever their origin, from zero-modes of the  symplectic matrix $f$, 
	from the equations of motion or even the {\it ad-hoc} ones --- 
	{\it must} be introduced into the kinetic sector of the Lagrangian through a velocity multiplier.
\item If $f$ is a nonsingular matrix, all constraints {\it should} be
	strongly set to zero in its respective first-order Lagrangian, which
	includes the symplectic potential.
\end{itemize}

In order to clarify this proposal, the three models studied 
in section \ref{ComparisonDiracBW}
 will be revisited in the following subsections.

\subsection{Revisiting the Free Particle on a Hypersphere}
\label{HypersphereBW+}

The dynamics of a {\it free} particle is governed by the Lagrangian
\ba 
\label{0320}
{\cal L}_{\text{hyp}}(\dot q)&=& \frac 12 \sum_{i=1}^{N}\dot q_i^2,
\ea
which can be written as a linear function of velocities as
\ba 
\label{0320a}
{\cal L}_{\text{hyp}}(\dot q,p)&=&\sum_{i=1}^{N}p_i\dot q_i - \frac 12 \sum_{i=1}^{N} p_i^2,
\ea
upon the introduction of the variables $p_i$'s.
By imposing the {\it ad-hoc} constraint
\be \label{0320b}
\Omega_1=\frac 12\left(\sum_{i=1}^{N}q_i^2-1\right),
\ee
the particle dynamics is restricted to be on a hypersphere of unit radius.
As the BW formalism starts out from
the first-order Lagrangian (\ref{0320a}), we introduce the constraint (\ref{0320b})
{\it in its kinetic sector} through a velocity multiplier, obtaining
\be 
\label{0330}
{\cal L}_{\text{hyp}}^{(1)}(\dot q,p,\dot \eta_1)= \sum_{i=1}^{N} p_i\dot q_i +\Omega_1\dot\eta_1- \frac12\sum_{i=1}^{N} p_i^2.
\ee 
This theory belongs to class of theories where part of the
coordinates does not have any time derivatives in the Lagrange function, so they are called degenerate coordinates \cite{gitman}. In the generalized Hamiltonization procedure the degenerate coordinates does not complete with the corresponding conjugate momenta. The authors of ref.\cite{gitman} demonstrated that the degenerate coordinates may be treated on the same footing as usual velocities. In fact, in the Hamiltonization procedure of the Maxwell theory, $A_0$ is considered a Lagrange multiplier to
a constraint and not a conjugate momentum to $A_0$ is introduced \cite{dirac4}.

Note that our proposition is not the same as the one found in
Refs.\cite{BW1,BW2,Shirzad1,rothe1,Shirzad2,rothe2}.
The symplectic variables now form the vector
\be
\xi^{(1)}=\left(
\begin{array}{c}
\q     \cr
\p     \cr
\eta_1 
\end{array}\right),
\ee
and the respective $f_{\text{hyp}}^{(1)}$ matrix reads
\be
\label{0340}
f_{\text{hyp}}^{(1)}=\left(
\begin{array}{ccc}
\O    ~&~  -\I  ~&~ ~\q \cr
\I    ~&~  \O   ~&~ ~\o \cr
-\q^T ~&~  \o^T ~&~ ~0
\end{array}\right).
\ee
This matrix is singular; consequently, it has a zero-mode, namely
\be
\nu^{(1)} =\left(
\begin{array}{c}
\o \cr
\q \cr
1  
\end{array}\right),
\ee
which generates a new constraint, equivalent to (\ref{0150}),
which is then inserted into the kinetic sector of (\ref{0330}).
We obtain
\be 
\label{0341}
{\cal L}_{\text{hyp}}^{(2)}(\dot q,p,\dot \lambda)= \sum_{i=1}^{N} p_i\dot q_i +\Omega_1\dot\eta_1+\Omega_2\dot\eta_2- \frac12\sum_{i=1}^{N} p_i^2.
\ee 
From (\ref{0341}) the nonsingular symplectic matrix (\ref{0170}), is restored. Therefore, all the constraints should be
strongly set to zero, which changes the first-order Lagrangian (\ref{0341}), to
\be 
\label{0342}
\left.{\cal L}^{(2)}_{\text{hyp}}\rule{0mm}{4mm}\right|_{\Omega_1=\Omega_2=0}= \sum_{i=1}^{N} p_i\dot q_i - \frac12\sum_{i=1}^{N} p_i^2,
\ee 
the canonical Legendre transformation, in which the symplectic
potential is the Hamiltonian (${\cal H}=\frac12\sum_{i=1}^{N} p_i^2)$.
Dirac's brackets among the phase-space coordinates are given by the (\ref{0040}). This procedure reproduces what has been obtained in
subsection \ref{sec21}, but the constraints should be set strongly to zero in the first-order Lagrangian, since its respective symplectic
matrix is nonsingular.

\subsection{Revisiting The Toy Model}
\label{ToyBW+}

In this subsection, we will continue the discussion carried out in
subsection \ref{sec22}, but considering that the constraint $\Theta_1$
has {\it not} been set strongly to zero.
Hence, the Lagrangian reads
\be 
\label{0350}
{\cal L}_{\text{toy}}^{(1)}(\dot x,\dot y,p_x,p_y,\dot\lambda_1)=p_x \dot x-ax\dot y + \Theta_1\dot\lambda_1- {\cal H}_{\text{toy}},
\ee
similar to Eq.(\ref{0230}) but with the important difference that ${\cal H}_{\text{toy}}$ 
is still given by (\ref{0190}), before any constraint is
imposed. 
The symplectic variables are
\be
\xi^{(1)}=\[x,p_x,y,\lambda_1\]
\ee
and the corresponding matrix $f_{\text{toy}}^{(1)}$ is given by (\ref{0250}), the inverse 
of which corresponds to (\ref{0270}). The Dirac brackets among the variables
are identified as the elements of such inverse matrix and Hamilton's equations are obtained,
\ba
\label{0360}
\dot x&=&\lbrace x,{\cal H}_{\text{toy}}\rbrace= \frac{b}{a^2-b}\left[\rule{0mm}{4mm}-p_x+a(x-y)\right],\nonumber\\
\dot p_x&=&\lbrace p_x,{\cal H}_{\text{toy}}\rbrace=0,\nonumber\\
\dot y&=&\lbrace y,{\cal H}_{\text{toy}}\rbrace=\frac{b}{a^2-b}\left[\rule{0mm}{4mm}-p_x+a(x-y)\right],\\
\dot \lambda_1&=&\lbrace \lambda_1,{\cal H}_{\text{toy}}\rbrace=\frac{\Theta_1}{(a^2-b)}.\nonumber
\ea
From the time derivative of the third equation in (\ref{0360}), we get
\be  
\label{0370}
\ddot y = \frac{ab}{a^2-b}(\dot x-\dot y) = 0,
\ee 
in which the first equation of (\ref{0360}) was also used. From
Eq.(\ref{0370}) we get that $\dot y = c,$ constant, and using the third
equation of (\ref{0360}), a new constraint is obtained,
\be
\label{0380}
\Gamma = \Theta_2 -c^\prime,
\ee
in which $\Theta_2$ is given by (\ref{0080}) and $c^\prime=(a^2-b)c$.

In agreement with the modified BW algorithm, the constraint (\ref{0380})
is then introduced into the kinetic sector of the Lagrangian
(\ref{0350}), yielding
\be 
\label{0390}
{\cal L}_{\text{toy}}^{(2)}(\dot x,\dot y,\dot\lambda_1,\dot\lambda_2,p_x,p_y)=p_x \dot x-ax\dot y + \Theta_1\dot\lambda_1+\Gamma\dot\lambda_2- {\cal H}_{\text{toy}}.
\ee
The symplectic variables are 
\be
\xi^{(2)}=\[x,p_x,y,\lambda_1,\lambda_2\]
\ee 
and the corresponding $f$ matrix reads
\be
\label{0400}
f_{\text{toy}}^{(2)}=\left( 
\begin{array}{ccccc}
0&-1&-a&-b&+ab\\
+1&0&0&+a&-b\\
+a&0&0&+b&-ab\\
+b&-a&-b&0&0\\
-ab&+b&+ab&0&0
\end{array}
\right),
\ee
which is singular. Consequently, it has a zero-mode of the form
\be
\label{0409}
\nu^{(2)}=\left(
\begin{array}{c}
-b\cr0\cr-b\cr0\cr-1
\end{array}
\right),
\ee
which, after the contraction with the gradient of the symplectic potential [still given by Eq.(\ref{0190})], yields
\be 
\label{0410}
\nu^{(2)T}\,\frac{\partial{\cal H}_{\text{toy}}}{\partial\xi^{(2)}}= 0.
\ee
No new constraint arises; consequently, the model has a symmetry and the zero-mode (\ref{0409}) is the generator of the infinitesimal symmetry transformation ($\delta\xi=\varepsilon\nu^{(2)}$), 
given in Eq.(\ref{0095}),
which keeps the Lagrangian (\ref{0390}) invariant.
The Hamiltonian (\ref{0190}) transforms as
\be 
\label{0419}
\delta {\cal H}_{\text{toy}}=0.
\ee
By following this approach, we obtain the same results as in Dirac's formalism, discussed in subsection \ref{sec12}.

\subsection{Revisiting the Relativistic free particle}
\label{subsec4c}
\label{RelativisticBW+}

In accordance with the proposed modifications to the BW formalism, stated at the beginning of section
\ref{sec3}, the constraint (\ref{00080}) must be inserted into the
first-order Lagrangian (\ref{000100}) in its kinectic sector. 
Hence\footnote{In this subsection \ref{RelativisticBW+}, we use the convention of repeated indices to represent summation.}, 
\be 
\label{000120}
{\cal{L}}=-p^\mu\dot x_\mu+\phi\dot\lambda.
\ee
The symplectic variables are $\xi^\alpha=(x_\mu, p^\mu,\lambda)$ and the symplectic matrix is
\be 
\label{000130}
f=\left(
\begin{array}{ccc}
0 & \delta^\mu_\nu & 0\\
-\delta^\mu_\nu & 0 & 2p^\nu\\
0 & -2p^\mu & 0
\end{array}
\right).
\ee
This matrix is singular and has a zero-mode, which reads
\be 
\label{000140}
\nu=\left(
\begin{array}{c}
2p^\nu\\
0\\
1
\end{array}
\right).
\ee
The contraction of this zero-mode 
with the gradient of the symplectic potential, $(V=0)$ is identically
null. Therefore, this system presents a symmetry and the zero-mode (\ref{000140})
 is the generator of the infinitesimal transformation
given by (\ref{00090}). 

In order to fix the symmetry, the constraint $\Sigma=x_0-ct$\footnote{As shown in Ref.\cite{gitman1}, section 7.3, this fixing condition is excessively rigid.} is
introduced into the kinetic sector of first-order Lagrangian,
(\ref{000120}), namely:
\be 
\label{000150}
{\cal{L}}=-p^\mu\dot x_\mu+\phi\dot\lambda+\Sigma\dot\eta.
\ee
The symplectic variables are $\xi=(x_\mu,p^\mu,\lambda,\eta)$ and the respective matrix is
\be 
\label{000160}
f=\left(
\begin{array}{cccc}
\textbf{0} & \delta^\mu_\nu & \textbf{0}&\delta^\mu_0\\
-\delta^\nu_\mu & \textbf{0} & 2p_\mu&\textbf{0} \\
\textbf{0} & -2p_\nu & \textbf{0}&\textbf{0}\\
-\delta^\nu_0 &\textbf{0}&\textbf{0}&\textbf{0}
\end{array}
\right).
\ee
This matrix is nonsingular and its inverse is
\be 
\label{000170}
f^{-1}=\left(
\begin{array}{cccc}
\textbf{0} & -\delta^\mu_\nu +\delta^\mu_0\frac{p_\nu}{p_0}&\textbf{0}&-\frac{p_\nu}{p_0}\\
\delta^\nu_\mu -\delta^\nu_0\frac{p_\mu}{p_0} & \textbf{0} & -\frac{\delta^\nu_0}{2p_0}& \textbf{0}\\
\textbf{0} & \frac{\delta^\mu_0}{2p_0} & \textbf{0}&-\frac{1}{2p_0}\\
\frac{p_\mu}{p_0}&\textbf{0}&\frac{1}{2p_0}&0
\end{array}
\right),
\ee
in which the non-null Dirac's brackets are identified as
\be 
\label{000180}
\lbrace p^\nu, x_\mu\rbrace= \delta^\nu_\mu -\delta^\nu_0\frac{p_\mu}{p_0}.
\ee
As the symplectic matrix is nonsingular, the constraints $\phi$ and
$\Sigma$ in the Lagrangian, given in Eq.(\ref{000150}), should be
strongly set to zero, then 
\ba 
\label{000190}
{\cal{L}}&=&-p^\mu\dot x_\mu+\phi\dot\lambda+\Sigma\dot\zeta,\nonumber\\
&=&-p^0\dot x_0-p^i\dot x_i+\phi\dot\lambda+\Sigma\dot\zeta,\\
&=&p_i\dot x_i-p^0c,\nonumber\\
&=&p_i\dot x_i-{\cal{H}},\nonumber
\ea
where the symplectic potential is the Hamiltonian given in Eq.(\ref{00070}) and the Legendre transformation is obtained as well. As $\phi$ is strongly set to zero, we get $p^0=\sqrt{p_i^2+m^2c^2}$, consequently, the Hamiltonian, given in Eq.(\ref{000190}), changes to
\be 
{\cal{H}}=c\sqrt{p_i^2+m^2c^2},
\ee
which is the well-known Hamiltonian for the relativistic free particle.

In order to put this resolution in a correct perspective to the general Faddeev-Jackiw symplectic approach to non-autonomous constrained systems\cite{Gavrilov}. The gauge fixing conditions is $(\Sigma=x_0-\zeta ct)$ and the variable $\zeta$ is not fixed by constraints and should be considered as an equal rights dynamical variable, which can assume two values $\zeta=\pm 1$. According to Ref.\cite{Gavrilov}, $\zeta$ should be also fixed, so $\Psi=\zeta-1$. Consequently and in agreement with Modified Barcelos-Wotzasek symplectic algorithm, the first-order Lagrangian, Eq.(\ref{000150}), changes to
\be 
\label{000150a}
{\cal{L}}=-p^\mu\dot x_\mu+\phi\dot\lambda+\Sigma\dot\eta+\Psi\dot\gamma.
\ee
The symplectic variables are $\xi=(x_\mu,p^\mu,\lambda,\eta,\zeta,\gamma)$ and the respective matrix is
\be 
\label{000160a}
f=\left(
\begin{array}{cccccc}
\textbf{0} & \delta^\mu_\nu & \textbf{0}&\delta^\mu_0&0&0\\
-\delta^\nu_\mu & \textbf{0} & 2p_\mu&\textbf{0} &0&0\\
\textbf{0} & -2p_\nu & \textbf{0}&\textbf{0}&0&0\\
-\delta^\nu_0 &\textbf{0}&\textbf{0}&\textbf{0}&0&0\\
0&0&0&0&0&1\\
0&0&0&0&-1&0
\end{array}
\right),
\ee
which after some straightforward computation restore the result previously obtained in this section. 
\section{General formulation of the Modified Barcelos-Wotzasek symplectic algorithm}

Consider a general n-dimensional unconstrained system whose its dynamics
is described by a Lagrangian 
\be
{\cal{L}}\equiv{\cal{L}}(x_i,\dot x_i),
\ee
in which $i\in\{1,2,\dots,n\}$. The Lagrangian depends
not only on the generalized coordinates $x_i$, but also on the
generalized velocities $\dot x_i$; hence, $\cal L$ is a function on a
manifold larger than the configuration space $\mathbb {Q}$, the
\emph{velocity phase manifold} (which which is not a vector space, but
the \emph{tangent bundle} or \emph{tangent manifold} \textbf{T}$\mathbb
{Q}$ of the configuration manifold $\mathbb {Q}$). 

Let us to explain this point in more detail. The \emph{tangent bundle}
\textbf{T}$\mathbb {Q}$ is obtained from $\mathbb {Q}$ by adjoining to
each point, $x_i\in\mathbb {Q}$, the tangent space
\textbf{T}$_{x_i}\mathbb {Q}$, which includes all possible velocities at
$x_i$, all tangent to $\mathbb {Q}$ at $x_i$. The generalized momentum
$p_i$, canonically conjugated to $x_i$, is
\be 
\label{000195}
\p_i=\frac{\partial {\cal{L}}(x_i,\dot x_i)}{\partial \dot x_i}.
\ee
It is possible, then, to rewrite the Lagrangian as a first-order function of the velocities,
\be 
\label{000200a}
{\cal{L}}^{(0)}=\sum_{i=1}^n p_i\dot x_i - V(x_i,p_i), 
\ee
in which $V(x_i,p_i)$ is called the {\it symplectic potential}. This
potential could be identified as the Hamiltonian, 
provided that the velocities $\dot x_i$ be written as functions of the momenta $p_i$, e.g., by the
inverse of Eq.(\ref{000195}). In the Faddeev-Jackiw
symplectic algorithm, the Lagrangian ${\cal{L}}(x_i,\dot x_i)$ is
rewritten in a first-order form, that is, as
\be 
\label{000200b}
{\cal{L}}^{(0)}= \sum_{i=1}^{2n} A_\alpha^{(0)}\dot \xi_\alpha^{(0)} -V(\xi_\alpha^{(0)}).
\ee
in which the symplectic variable and the generalized momentum
(usually called the one-form momentum) are, respectively, $\xi_\alpha^{(0)}=(x_i,p_i)$ and $A_\alpha^{(0)}=(p_i, 0_i)$, which belongs to the
\emph{cotangent bundle} or the \emph{phase manifold} (phase space)
\textbf{T}$^\ast\mathbb {Q}$. 
After that, the symplectic matrix
\be
f_{\xi_\alpha\xi_\beta}^{(0)}=\partial_{\xi_\alpha^{(0)}} A_\beta^{(0)}-\partial_{\xi_\beta^{(0)}} A_\alpha^{(0)}
\ee 
is computed. As $\det{f^{(0)}\neq 0}$, $f^{(0)}$ is inversible and $V(\xi_\alpha^{(0)})$ is identified as the Hamiltonian --- because it is possible to write the
velocities in terms of the momenta. Consequently, the pullback of the
Legendre transformation is obtained. Further, the Hamilton's equation of motion might be computed as well, which are
$\dot\xi^{(0)}_\alpha=\lbrace \xi^{(0)}_\alpha,
V(\xi^{(0)}_\alpha)\rbrace$, then $\mathbb {Q}\longrightarrow\textbf{T}\mathbb
{Q}\longrightarrow\textbf{T}^\ast\mathbb {Q}$.

At this point, a general n-dimensional constrained system with $m$ initial constraints arising
from the Lagrange framework -- either constraints of geometric nature
or (in Dirac's language) the primary constraints, is considered. In the modified Barcelos-Wotzasek
symplectic algorithm, the Lagrangian ${\cal{L}}(x_i,\dot x_i)$ is
rewritten in a first-order form and the $m$ constraints are
introduced into the kinetic sector, that is, as
\be 
\label{000200}
{\cal{L}}^{(0)}= \sum_{i=1}^n p_i\dot x_i +\sum_{a=1}^m \Omega_a(p_i,x_i)\dot\eta_a -V(x_i,p_i),
\ee
in which $a\in \{1,2,\dots,m\}$, $\Omega_a(p_i,x_i)$ is the $a$-th constraint and
$\dot\eta_a$ is the respective $a$-th velocity multiplier. Note that the
velocity $\dot x_i$ was {\it not} written in terms of $p_i$ and, indeed,
Eq.(\ref{000195}) might be a constraint. The first-order Lagrangian (\ref{000200}) is rewritten as
\be 
\label{000210}
{\cal{L}}^{(0)}=\sum_{\alpha=1}^{2n+m} A_\alpha^{(0)}\dot\xi_\alpha^{(0)}  -V(\xi_\alpha^{(0)}), 
\ee
in which the symplectic variable and the generalized momentum are, respectively, $\xi_\alpha^{(0)}=(x_i,p_i,\eta_a)$ and $A_\alpha^{(0)}=(p_i, 0_i,\Omega_a)$, which belongs to the expanded
\emph{cotangent bundle} \textbf{T}$^\ast\mathbb {Q}$. We would like to emphasize that  the  \textbf{T}$^\ast\mathbb {Q}$ is expanded until its dimension turns even. Of course, this implies that the configuration space $\mathbb {Q}$ is proportionally expanded in order to give room to the constraints inserted to the Lagrangian through velocity multipliers. After that, the symplectic matrix
\be
f_{\xi_\alpha\xi_\beta}^{(0)}=\partial_{\xi_\alpha^{(0)}} A_\beta^{(0)}-\partial_{\xi_\beta^{(0)}} A_\alpha^{(0)}
\ee 
is computed. If $2n+m$ is odd, then $f^{(0)}$ is singular, which means the existence of a new constraint or a symmetry. On the other hand, $(2n+m)$ is even, $f^{(0)}$ can be still singular or can be nonsingular, which means that there are no more constraints and the generalized brackets among the
symplectic variables are obtained from elements of the inverse of
$f^{(0)}$. At this point, the $m$ constraints must
be set to zero in the first-order Lagrangian,
\be 
\label{000220}
\left.{\cal{L}}^{(0)}\rule{0mm}{0.7mm}\right|_{\Omega=0}=\left.\left[\sum_{\alpha=1}^{2n+m}A_\alpha^{(0)}\dot\xi_\alpha^{(0)}  -V(\xi_\alpha^{(0)})\right]\rule{0mm}{0.7mm}\right|_{\Omega=0},
\ee
in which 
$\Omega=0$ stands for the set of $m$ equations $\{\Omega_a = 0 \ | \  a \in \{1,2,\dots, m\}\}$, and 
$\left.V(\xi_\alpha^{(0)})\rule{0mm}{0.7mm}\right|_{\Omega=0}$ is
identified as the Hamiltonian --- because it is possible to write the
velocities in terms of the momenta. Consequently, the pullback of the
Legendre transformation is obtained, which allows one to get the
Lagrangian after the constraints were set to zero. Simultaneously,
Hamilton's equation of motion might be computed as well, which are
$\dot\xi^{(0)}_\alpha=\lbrace \xi^{(0)}_\alpha,
\left.V(\xi_\alpha^{(0)})\rule{0mm}{0.7mm}\right|_{\Omega=0}\rbrace$. At this
point, $\mathbb {Q}\longrightarrow\textbf{T}\mathbb
{Q}\longrightarrow\textbf{T}^\ast\mathbb {Q}$.

On the other hand, if the symplectic matrix
$f_{\xi_\alpha\xi_\beta}^{(0)}$ is singular, it has a zero-mode
that generates a new constraint $\Sigma_1$ when contracted with the
gradient of the symplectic potential. This constraint must be inserted
into the kinetic sector of the first-order Lagrangian (\ref{000210})
through a velocity multiplier $\dot\zeta_1$,
\be 
\label{000230}
{\cal{L}}^{(1)}=\sum_{i=1}^n p_i\dot x_i +\sum_{a=1}^m \Omega_a\dot\eta_a +\Sigma_1\dot\zeta_1-V(\xi_\alpha^{(1)}).
\ee
The new symplectic variables are
$\xi_\alpha^{(1)}=(x_i,p_i,\eta_a,\zeta_1)$ and the new symplectic
matrix 
$f_{\xi_\alpha\xi_\beta}^{(1)}$
is computed,
\be
f_{\xi_\alpha\xi_\beta}^{(1)}=\partial_{\xi_\alpha^{(1)}} A_\beta^{(1)}-\partial_{\xi_\beta^{(1)}} A_\alpha^{(1)}.
\ee

In its turn, if $f_{\xi_\alpha\xi_\beta}^{(1)}$ is nonsingular, then the
generalized brackets among the new symplectic variables are obtained and
all constraints must be set to zero in the new Lagrangian,
\be 
\label{000240}
\left.{\cal{L}}^{(1)}\rule{0mm}{0.7mm}\right|_{\tiny{\left(\begin{array}{ll}
\Omega=0\\
\Sigma_1=0
\end{array}\right)}}=\left.\left[\sum_{\alpha=1}^{2n+m+1} A_\alpha^{(1)}\dot\xi_\alpha^{(1)}  -V(\xi_\alpha^{(1)})\right]\rule{0mm}{0.7mm}\right|_{\tiny{\left(\begin{array}{ll}
\Omega=0\\
\Sigma_1=0
\end{array}\right)}},
\ee
where $(2n+m+1)$ is even. The symplectic potential, given by
\be 
\left.V(\xi_\alpha^{(1)})\rule{0mm}{0.7mm}\right|_{\tiny{\left(\begin{array}{ll}
\Omega=0\\
\Sigma_1=0
\end{array}\right)}}
\ee
can then be identified as the Hamiltonian and, after all constraints were 
set to zero, the Lagrangian is obtained by the pullback of Legendre
transformation, the Hamilton's equation of motion might be computed,
$\dot\xi^{(1)}_\alpha=\lbrace \xi^{(1)}_\alpha,
\left.V(\xi_\alpha^{(1)})\rule{0mm}{0.7mm}\right|_{\Omega=0, \Sigma_1 = 0}\rbrace$ and
\textbf{T}$\mathbb {Q}$ is mapped to \textbf{T}$^\ast\mathbb {Q}$.

However, if $f_{\xi_\alpha\xi_\beta}^{(1)}$ is singular, then there will
be a new zero-mode which generates a new constraint $\Sigma_2$ and so
on. After $k$ interactions, two distinct possible scenarios may emerge. 
\begin{enumerate}
\item The symplectic matrix
	$f_{\alpha\beta}^{(k)}$ is still singular, 
	but its zero-mode does not generate a new constraint (or generates a previous one).
	The system than has a symmetry, and the zero-mode is its infinitesimal generator.
	This symmetry could be fixed (a new constraint!) as in the BW formalism.

\item The
symplectic matrix $f_{\alpha\beta}^{(k)}$ is nonsingular. Then the
generalized brackets among the symplectic variables $\xi_\alpha^{(k)}$
are obtained from the inverse of the symplectic matrix $f_{\alpha\beta}^{(k)}$ and the
first-order Lagrangian reads
\be 
	\label{000250}
	\left.{\cal{L}}^{(k)}\rule{0mm}{0.7mm}\right|_{\tiny{\left(\begin{array}{ll}
	\Omega=0\\
	\Sigma=0
	\end{array}\right)}}=\left.\left[\sum_{\alpha=1}^{2n+m+k} A_\alpha^{(k)}\dot\xi_\alpha^{(k)}  -V(\xi_\alpha^{(k)} )\right]\rule{0mm}{0.7mm}\right|_{\tiny{\left(\begin{array}{ll}
	\Omega=0\\
	\Sigma=0
	\end{array}\right)}},
	\ee
in which $(2n+m+k)$ is even, 
$\Sigma=0$ stands for the set of $k$ equations $\{\Sigma_j = 0 \ | \  j \in \{1,2,\dots, k\}\}$, and 
$\left.V(\xi_\alpha^{(k)} )\rule{0mm}{0.7mm}\right|_{\tiny{\left(\begin{array}{ll}
\Omega=0\\
\Sigma=0
\end{array}\right)}}$ 
is identified as the Hamiltonian. 
Consequently, the pullback of the Legendre transformation is obtained,
which allows one to get the Lagrangian after all constraints are set to zero. 
Furthermore, Hamilton's equations of motion are also
obtained, $\dot\xi^{(k)}_\alpha=\lbrace \xi^{(k)}_\alpha,
\left.V(\xi_\alpha^{(k)} )\rule{0mm}{0.7mm}\right|_{\Omega=0, \Sigma=0}\rbrace$. This
procedure maps the configuration space to the phase space and
\textit{vice-versa} for constrained systems, e.g., this procedure
restores the Legendre transformation for constrained systems, as well
illustrated in sections \ref{HypersphereBW+} and \ref{RelativisticBW+}.
\end{enumerate}
\section{Conclusion}
\label{conclusion}

There has been some criticism on the BW symplectic formalism. Indeed, 
such formalism may return the same results of Dirac's method for some
models, as exemplified in subsection \ref{HypersphereDiracBW}. There
are other models, however, for which the results attained by the BW
formalism may disagree with those of Dirac's, as exemplified in
subsections \ref{ToyDiracBW} and \ref{RelativisticDiracBW}. 

We propose a minor modification in the BW formalism to account for those
disagreements. Namely, that each and every constraint should be introduced
into the kinetic sector through a velocity
multiplier, and these constraints should {\it not} be strongly set to zero
if the $f$ matrix is singular. On the other hand, if the $f$ matrix
is nonsingular, the constraints inserted into the kinetic sector should
indeed be strongly set to zero in the first-order Lagrangian, as shown
in the subsections of section \ref{BW+}. Therefore, this
modified BW symplectic algorithm turns out to be completely
equivalent to the well-established Dirac's method in its application to
constrained models.

Further, the Modified Barcelos-Wotzasek symplectic algorithm embraces the general Faddeev-Jackiw symplectic approach to non-autonomous constrained systems\cite{Gavrilov}, as shown in section \ref{RelativisticBW+}.

\section*{Authors' Declarations}

\begin{itemize}
\item \textbf{Ethical conduct:} this manuscript complies with the ethical policies of the journal.
\item \textbf{Competing interests:} the authors have no competing interests that might influence the results and/or discussion reported in this paper.
\item \textbf{Data avaliability:} all of the material is owned by the authors and/or no permissions have been required. No data sets have been generated during the current study.
\item \textbf{Funding:} This work is partially supported by FAPERJ and CNPq (Brazilian Research Agencies).
	E.V. Corr\^{e}a Silva thanks Universidade do Estado do Rio de Janeiro (UERJ) for the Proci\^{e}ncia grant.
\end{itemize}

%
%

\end{document}